\documentclass[12 pt]{article}    

\usepackage{pifont}
\usepackage{alltt}
\usepackage{graphicx}
\usepackage{caption2}
\usepackage{times}
\usepackage{tipa}
\usepackage{indentfirst}
\usepackage{dcolumn}%
\usepackage{amsmath,amsfonts,bm}%
\usepackage{amssymb}
\usepackage{color}

\newcounter{saveeqn}

\usepackage{fancyhdr}
\begin{document}

\begin{center}
\vskip 2 in 
\Large{\bf Simultaneous estimation of attenuation and structure parameters of aggregated red blood cells from backscatter measurements}\\ 
\vskip 1in
\large{Emilie Franceschini, Fran\c cois T. H. Yu, Guy Cloutier}\\ 
\vskip .1 in
\normalsize{Laboratory of Biorheology and Medical Ultrasonics}   \\
\vskip .1 in
\normalsize{University of Montreal Hospital Research Center}   \\
\vskip .1 in
\normalsize{2099 Alexandre de S\`eve (room Y-1619),}   \\
\vskip .1 in
\normalsize{Montr\'eal, Qu\'ebec, H2L 2W5, Canada}   \\
\vskip .1 in
\normalsize{Emilie.Franceschini@crchum.qc.ca}   \\
\vskip .1 in
\normalsize{Guy.Cloutier@umontreal.ca}   \\
\end{center}

\newpage
\baselineskip 28 pt

\section*{\normalsize Abstract}

The analysis of the ultrasonic frequency-dependent backscatter coefficient of aggregating red blood cells reveals information about blood structural properties. The difficulty to apply this technique \emph{in vivo} is due to the frequency-dependent attenuation caused by intervening tissue layers that distorts the spectral content of backscattering properties from blood microstructures. An optimization method is proposed to simultaneously estimate tissue attenuation and blood structure factor. In an \emph{in vitro} experiment, the method gave satisfactory estimates with relative errors below 22$\%$ for attenuations between 0.101 and 0.317 dB/cm/MHz, signal-to-noise ratios$>$28 dB and $kR<$2.7 ($k$ being the wave number and $R$ the aggregate radius).

\vskip .5cm
\noindent PACS numbers: 43.80.Qf, 43.80.Cs, 43.35.Bf, 43.35.Yb

\newpage

\section*{\normalsize 1. Introduction}

Ultrasonic (US) backscattered echoes from blood contain frequency-dependent information that can be used to obtain quantitative parameters reflecting the aggregation state of red blood cells (RBCs). Recently, two parameters describing RBC aggregation, the packing factor and mean fractal aggregate diameter, were extracted from the Structure Factor Size Estimator (SFSE).$^1$ The SFSE is a second-order data reduction model based on the structure factor and adapted to a dense medium such as blood. This approach is based on the analysis of the backscattered power spectrum that contains information about the size, spatial organization, concentration and mechanical properties of scatterers (i.e. RBCs). The difficulty to apply the SFSE \emph{in vivo} is that the spectral content of backscattered echoes is also affected by attenuation caused by intervening tissue layers between the probe and the blood flow. More generally, ultrasound scatterer size estimation techniques for tissue characterization (such as liver, kidney, prostate or breast) are facing similar challenges.$^{2,3}$ To evaluate correctly microstructural parameters, it is thus of major interest to take into account tissue attenuation effects. A few groups$^{2,4,5}$ developed measurement techniques to evaluate the frequency-dependent attenuation in order to compensate \emph{a posteriori} the backscattered power spectrum. The goal of this letter is to further develop this strategy for \emph{in vivo} measures of RBC scatterer sizes. We propose to determine simultaneously blood structural parameters and total attenuation by using an optimization method, termed the Structure Factor Size and Attenuation Estimator (SFSAE).

This method consists in fitting the spectrum of the backscattered radio-frequency (rf) echoes from blood to an estimated spectrum by a modified SFSE model. This approach is similar to that presented by Bigelow \emph{et al.}$^3$, who estimated the effective radius of tissue microstructure and total attenuation from simulated backscattered signals. Herein, \emph{in vitro} experimental evaluation of the SFSAE is performed. Porcine RBCs were sheared in a Couette flow system, and ultrasonic rf echoes were obtained using a 25 MHz center-frequency transducer. Since skin is one of the most attenuating tissue layers during \emph{in vivo} scanning, three skin-mimicking phantoms with different attenuation coefficients were successively introduced between the transducer and the blood flow. This study shows the SFSAE ability to evaluate three parameters (the packing factor, mean fractal aggregate diameter and total attenuation).

\section*{\normalsize 2. Structure Factor Size and Attenuation Estimator}

Blood can be seen as a very dense suspension of red cells. These RBCs cannot be treated as independent scatterers since particle interactions (collision, attraction, deformation, flow dependent motions) are strong. The theoretical model of ultrasound backscattering by blood that we developped$^1$ is based on the particle approach$^{6,7}$, which consists of summing contributions from individual RBCs and modeling the RBC interaction by a particle pair-correlation function. Assuming that the Born approximation is valid (weak scattering), the model proposed in Ref. 1 can be modified to predict the theoretical backscatter coefficient from blood:
\begin{equation}\label{eq:BSCwithS}
BSC_{theor}(k)=m\sigma_b(k)S(k)A(k)
\end{equation}
where $k$ is the wave vector, $m$ is the number density of RBCs in blood estimated by measuring the hematocrit $H$ by microcentrifugation ($m=H/V_s$, where $V_s$ is the volume of a RBC), $\sigma_b$ is the backscattering cross section of a single RBC, $S$ is the structure factor describing the spatial organization of RBCs, and $A$ is the frequency-dependent attenuation function. The backscattering cross-section $\sigma_b$ of a weak scattering particle small compared to the wavelength (Rayleigh scatterer) can be determined analytically as follows: $\sigma_b(k)=\displaystyle 1/(4\pi^2) k^4 V_s^2 \gamma_z^2$, where $\gamma_z$ is the variation of impedance between the RBC and its suspending medium (i.e. the plasma). The structure factor $S$ is by definition the Fourier transform of the pair-correlation function$^7$ $g$ and can be approximated by its second-order Taylor expansion$^1$ in $k$ as
\begin{equation}\label{eq:defS}
S(k)=1+m\displaystyle \int \left(g(r)-1 \right) e^{-2jkr} dr\approx W-\frac{12}{5}(kR)^2.
\end{equation} 
In this expression, $g(r)$ represents the probability of finding two particles separated by a distance $r$. $W$ is the low-frequency limit of the structure factor ($S(k)|_{k\rightarrow 0}$) called the packing factor.$^{7,8}$ $R$ is the radius of 3D RBC aggregates assumed to be isotropic. We introduce $D=R/a$ as the isotropic diameter of an aggregate (expressed in number of RBCs) with $a$ the radius of one RBC sphere-shaped model of volume $V_s$. The attenuation function $A$ is given by: $A(k)=\displaystyle e^{-4 \alpha_0 f}$, where $f$ is the frequency and $\alpha_0$ is the attenuation coefficient (in dB/MHz) defined by: $\alpha_0 = \displaystyle \sum_i \alpha_{i} e_{i}$, where $\alpha_{i}$ and $e_{i}$ are respectively the intervening tissue layer attenuations (in dB/cm/MHz) and thicknesses. According to the above equation, we thus assume, as a first approximation, that the attenuation increases linearly with the frequency: $\alpha (f)=\alpha_0 f$.

The measured backscatter coefficient reported in this study was computed as
\begin{equation}\label{eq:BSCmeas}
BSC_{meas}(k)=BSC_{ref}(k) \displaystyle \frac{\overline{P_{meas}(k)}}{\overline{P_{ref}(k)}}.
\end{equation}
In Eq. (3), the mean backscattered power spectrum $\overline{P_{meas}}$ was obtained by averaging the power spectra of 400 backscattered echoes from blood. The mean power spectrum $\overline{P_{ref}}$ was obtained from a reference sample of non-aggregated RBCs at a low hematocrit of 6$\%$ (i.e. Rayleigh scatterers).$^9$ In this case, 400 echoes were also averaged. The backscatter coefficient of this reference sample $BSC_{ref}$ was estimated using the "Rayleigh estimation" approach used by Yu and Cloutier$^1$, which theoretical value is given by Eq. (13) in Ref. 8 (three dimensional Perkus Yevick packing factor for cylinders). This reference sample was used to compensate the backscattered power spectrum $\overline{P_{meas}}$ for the electromechanical system response, and the depth-dependent diffraction and focusing effects caused by the US beam.

The packing factor $W$, aggregate diameter $D$ and total attenuation along the propagation path $\alpha_0$ were determined by matching the measured $BSC_{meas}$ given by Eq. (\ref{eq:BSCmeas}) with the theoretical $BSC_{theor}$ given by Eq. (\ref{eq:BSCwithS}). For this purpose, we searched values of $W$, $D$ and $\alpha_0$ minimizing the cost function $F(W,D,\alpha_0)$=$||BSC_{meas}-BSC_{theor}||^2$. In all studied cases, the cost function seemed to have a unique global minimum, as was observed by plotting the cost function surface $F(W,D)$ with varying values of $\alpha_0$. An example is given in Fig. 1.

\section*{\normalsize 3. Methods}
\subsection*{\normalsize \emph{3.1 Experimental set up}}
US measurements were performed in a Couette flow system to produce a linear blood velocity gradient at a given shear rate (see figure 1 in Ref. 11). The system consisted of a rotating inner cylinder with a diameter of 160 mm surrounded by a fixed concentric cylinder of diameter 164 mm. A 60 mL blood sample was sheared in the 2 mm annular space between the two coaxial cylinders. The US scanner (Vevo 770, Visualsonics, Toronto, Canada) equipped with the RMV 710 probe was used in M-mode for this study. The single-element focused circular transducer had a center frequency of 25 MHz, a diameter of 7.1 mm and a focal depth of 15 mm. We operated at a sampling frequency of 250 MHz with 8 bits resolution (Gagescope, model 8500CS, Montreal, Canada). The probe was mounted in the side wall of the fixed outer cylinder and was positioned to have its focal zone at the center of both cylinders. To ensure ultrasonic coupling, the hole within the outer stationary cylinder (containing the probe) was filled with a liquid agar gel based mixture. When solidified, this gel was cut to match the curvature of the cylinder to avoid any flow disturbance.  The gel was a mixture of distilled water, 3$\%$ (w/w) agar powder (A9799, Sigma Chemical, Saint-Louis, MO), 8$\%$ (w/w) glycerol and a specific concentration of 50 $\mu m$ cellulose scattering particles (S5504 Sigmacell, Sigma Chemical, Saint-Louis, MO) that determined the attenuation coefficient. Four experiments were performed with four mixtures having Sigmacell (SC) concentrations varying from 0$\%$ to 0.75$\%$ (w/w). The 0$\%$ concentration constituted the non-attenuating gel and the three other mixtures mimicked skin attenuations.

\subsection*{\normalsize \emph{3.2 Attenuation measurements}}
The attenuation coefficients of the reference (0$\%$) and three skin-mimicking phantoms $\alpha_p$ were determined by using a standard substitution method. Two transducers with center frequencies of 25 MHz (Vevo 770, Visualsonics, Toronto, Canada) and 20 MHz (V317-SM Panametrics, Waltham, MA) were aligned coaxially facing each other for transmission measurements. Transmitted signals were recorded both with and without the agar gel sample in the acoustic path. The attenuation coefficient was then estimated using a log spectral difference technique.$^{10}$ For a given concentration of SC, measurements were obtained from two different sample thicknesses, and for each, four regions were scanned for averaging purpose. Values obtained were 0.007 $\pm$ 0.002, 0.101 $\pm$ 0.028, 0.208 $\pm$ 0.029 and 0.317 $\pm$ 0.039 dB/cm/MHz for SC concentrations of 0, 0.25, 0.50 and 0.75$\%$, respectively. The thickness of the skin-mimicking phantoms $e_p$ being fixed to 1 cm, their attenuation coefficients were thus in the same range as the human dermis (0.21 dB/MHz at 14 - 50 MHz considering a 1 mm dermis thickness$^{12}$).

\subsection*{\normalsize \emph{3.3 Blood preparation and measurement protocol}}
Fresh porcine whole blood was obtained from a local slaughter house, centrifuged and the plasma and buffy coat were removed. Two blood samples were then prepared: (i) a H6 reference sample, which was a 6$\%$ hematocrit non-aggregating RBCs resuspended in saline solution; and (ii) a 40$\%$ hematocrit T40 test sample, which consisted of RBCs resuspended in plasma to promote aggregation. The H6 sample was sheared at 50 s$^{-1}$ and coupled with the 0$\%$ SC concentration agar gel. Echoes were selected with a rectangular window of length 0.8 mm at four depths every 0.2 mm (i.e. with 75$\%$ overlap between windows). For each depth, the power spectra of the backscattered echoes were averaged over 400 acquisitions to provide $\overline{P_{ref}}$. Then, the H6 sample was removed and the T40 blood was introduced in the Couette device. In the first 30 s, a shear rate of 500 s$^{-1}$ was applied to disrupt RBC aggregates. The shear rate was then reduced to residual values of 2, 5, 10, 20 and 30 s$^{-1}$ for 90 s. After that, for each shear rate, acquisitions of 400 rf lines were performed for 80 s. Echoes were windowed as for the H6 sample at the same depths and their power spectra were averaged to obtain $\overline{P_{meas}}$. This protocol was repeated four times with the four agar-based phantoms.

\subsection*{\normalsize \emph{3.4 Reference measurements with the 0$\%$ SC concentration phantom}}
The experiment with the 0$\%$ SC phantom was realized in order to have reference results on packing factors $W_{ref}$ and aggregate diameters $D_{ref}$ obtained from the classical SFSE.$^1$ These parameters were assumed to be true values of packing factors and aggregate diameters for all shear rates, and will be compared in the next section with packing factors and diameters estimated by the SFSAE and by the SFSE when skin-mimicking phantoms are used.

It is important to emphasize the fact that the H6 reference sample was also measured with the 0$\%$ SC phantom. The phantom attenuation, although small with no SC, therefore affected equivalently both spectra $\overline{P_{meas}}$ and $\overline{P_{ref}}$ in Eq. (3). The resulting measured backscatter coefficient $BSC_{ref}$ was thus not biased by attenuation. The terminology ``no attenuation" was used for this experiment in the following.

\section*{\normalsize 4. Results and discussion}

Figure 2a reports results on $W_{ref}$ and $D_{ref}$ for the SFSE in the case of no attenuation. Typical results of the SFSAE minimization procedure for the different agar phantoms at a shear rate of 5 s$^{-1}$ are given in Fig. 2b. All results on $W$, $D$ and $\alpha_0$ from the SFSAE are summarized in Fig. 3 for all residual shear rates. In this figure, the relative errors for each parameter correspond to: $(W-W_{ref})/W_{ref}$, $(D-D_{ref})/D_{ref}$ and $(\alpha_0-\alpha_{ref})/\alpha_{ref}$, with $\alpha_{ref}$ measured
in transmissions. More specifically, $\alpha_{ref}$ corresponds to $\displaystyle \sum_i \alpha_{i} e_{i}=(\alpha_{p}e_{p}+\alpha_{blood}e_{blood})$, where $\alpha_pe_{p}$ is the skin-mimicking phantom attenuation estimated in transmission, and $\alpha_{blood}e_{blood}$ is the blood attenuation taken equal to 0.022 dB/MHz$^1$ for all shear rates. To underline the necessity to take into account the attenuation, parameters $W_{nocomp}$ and $D_{nocomp}$ were evaluated with the SFSE without attenuation-compensation when skin-mimicking phantoms were used. Because of the frequency-dependent distortion produced by the attenuating medium, large relative errors can be seen in Fig. 4a for both parameters. However, by compensating the backscatter coefficients in the SFSE with the value measured in transmission (section 3.2), relative errors in Fig. 4b are largely reduced to values comparable to those estimated with the SFSAE (see Fig. 3b).

The SFSAE (Fig. 3) gave quantitatively satisfactory estimates of $W$, $D$ and $\alpha_0$ with relative errors below 22$\%$, for shear rates between 5 and 20 s$^{-1}$. The SFSE with attenuation-compensation (Fig. 4b) gave estimates of $W_{comp}$ and $D_{comp}$ with relative errors below 12$\%$ for shear rates between 2 and 10 s$^{-1}$, and below 28$\%$ for the shear rate of 20 s$^{-1}$. However, for the SFSAE, the average estimates for the shear rate of 2 s$^{-1}$ were less accurate (relative errors below 57$\%$ for $W$ and below 30$\%$ for $\alpha_0$). The estimation of $D$ was satisfactory at that shear rate (relative errors below 14$\%$). The worse results of $W$, $D$ and $\alpha_0$ were obtained at 30 s$^{-1}$ for the highest attenuation.

The apparent limit of applicability of the SFSAE method for shear rates of 2 and 30 s$^{-1}$ may be explained by considering the following. At 2 s$^{-1}$, for the frequency bandwidth considered (9 - 30 MHz), the SFSE and consequently the SFSAE seem to reach their limit of applicability for large aggregate sizes (typically $D_{ref} =$ 17.5 in Fig. 2a, i.e. $kR$ = 4.8). This limit is illustrated by the bad fit of the SFSE model in Fig. 2a at 2 s$^{-1}$. The bad estimations of the SFSAE at 30 s$^{-1}$ are explained by the fact that the aggregate diameters were estimated to zero and attenuations were overestimated. At this high shear rate, RBC aggregation is partially inhibited and the signal-to-noise ratio (SNR) of our measurements was reduced ($\approx$ -4 dB between 20 and 30 s$^{-1}$ for all phantoms). The accuracy of the estimates was thus degraded with increasing attenuations, as can be seen from the large relative errors at the highest attenuation with the SFSAE but also with the SFSE with attenuation-compensation ($W_{comp}$ and $D_{comp}$).
 
To conclude, the SFSAE performed well for $kR <$ 2.7 (i.e. $D =$ 10 at 5 s$^{-1}$) and under the condition that the SNR is sufficiently good (SNR $>$ 28 dB corresponding to the SNR at 30 s$^{-1}$ for the 0.25$\%$ SC). Although the SFSAE gave less accurate estimates for 2 and 30 s$^{-1}$, the estimated parameter values presented in Fig. 3a show that the SFSAE gave qualitatively satisfactory estimates for the three SC skin-mimicking phantoms at all shear rates, since the estimates of $W$ and $D$ versus shear rates had the same behaviors as $W_{ref}$ and $D_{ref}$.

\section*{\normalsize 5. Conclusions}
The performance of the new SFSAE was assessed with experimental measurements on blood in a Couette flow device. The accuracy of the estimates obtained with the SFSAE was not as satisfactory as those obtained with the SFSE with attenuation-compensation (i.e when \emph{a priori} are known about the attenuation). Nevertheless, the SFSAE has the major advantage to be easily applicable \emph{in vivo} because of the simultaneous estimation of the blood structural properties and total attenuation (contrary to the SFSE attenuation-compensation method, needing the attenuation and thickness of the tissue intervening layers to be known). This work thus confirms the \emph{in vivo} applicability of RBC aggregate size and structure estimations. Complementary studies are nevertheless required to determine the validity domain of the SFSAE according to $kR$ and attenuation.

\section*{\normalsize Acknowledgments}
This work was supported by the Canadian Institutes of Health Research (grants $\sharp$MOP-84358 and CMI-72323), by the Heart and Stroke Foundation of Canada (grant $\sharp$PG-05-0313), and by the National Institutes of Health of USA (grant $\sharp$RO1HL078655). Dr Cloutier is recipient of a National Scientist award of the Fonds de la Recherche en Sant\'e du Qu\'ebec. We are also thankful to Dr F. Destrempes for his helpful discussion on the optimization tool.

\section*{\normalsize References and links}
\begin{itemize}

\item [] $^1$F. T. H. Yu and G. Cloutier, ``Experimental ultrasound characterization of red blood cell aggregation using the structure factor size estimator", J. Acoust. Soc. Am. \textbf{122}, 645-656 (2007). 

\item [] $^2$V. Roberjot, S. L. Bridal, P. Laugier, and G. Berger, ``Absolute backscatter coefficient over a wide range of frequencies in a tissue-mimicking phantom containing two populations of scatterers", IEEE Trans. Ultras., Ferroelect., Freq. Contr. \textbf{43}, 970-978 (1996).

\item [] $^3$T. A. Bigelow, M. L. Oelze, and W. D. O'Brien, ``Estimation of total attenuation and scatterer size from backscatter ultrasound waveforms", J. Acoust. Soc. Am. \textbf{117}, 1431-1439 (2005). 

\item [] $^4$P. He and J. F. Greenleaf, ``Application of stochastic analysis to ultrasonic echoes - Estimation of attenuation and tissue heterogeneity from peaks of echo envelope", J. Acoust. Soc. Am. \textbf{79}, 526-534 (1986). 

\item [] $^5$B. J. Oosterveld, J. M. Thijssen, P. C. Hartman, R. L. Romijn, and G. J. E. Rosenbusch, ``Ultrasound attenuation and texture analysis of diffuse liver disease: methods and preliminary results", Phys. Med. Biol \textbf{36}, 1039-1064 (1991).

\item [] $^6$L. Y. L. Mo and R. S. C. Cobbold, ``Theoretical models of ultrasonic scattering in blood", in \emph{Ultrasonic Scattering in Biological Tissues}, edited by K. K. Shung and G. A. Thieme (CRC, Boca Raton, FL, 1993), Chap. 5, pp. 125-170.

\item [] $^7$V. Twersky, ``Low-frequency scattering by correlated distributions of randomly oriented particles", J. Acoust. Soc. Am. \textbf{81}, 1609-1618 (1987).

\item [] $^8$K. K. Shung, ``On the ultrasound scattering from blood as a function of hematocrit", IEEE Trans. Ultras., Ferroelect., Freq. Contr. \textbf{SU-26}, 327-331 (1982).

\item [] $^9$S. H. Wang and K. K. Shung, ``An approach for measuring ultrasonic backscattering from biological tissues with focused transducers", IEEE Trans. Biomed. Eng. \textbf{44}, 549-554 (1997).

\item [] $^{10}$ R. Kuc and M. Schwartz, ``Estimating the acoustic attenuation coefficient slope for liver from reflected ultrasound signals", IEEE Trans. Sonics Ultrasonics \textbf{SU-26}, 353-362 (1979).

\item [] $^{11}$ L.-C. Nguyen, F. Yu, and Guy Cloutier, ``\emph{In Vitro} Study of Frequency-Dependent Blood Echogenicity under Pulsatile Flow", Proc.-IEEE Ultrason. Symp. 2007, in press.

\item [] $^{12}$B. I. Raju and M. A. Srinivasan, ``High-frequency ultrasonic attenuation and backscatter coefficients of \emph{in vivo} normal human dermis and subcutaneous fat", Ultrasound in Med. Biol. \textbf{27}, 1543-1556 (2001).

\end{itemize}

\newpage
\section*{\normalsize Figure captions}

\begin{itemize}
\item[] Figure 1. (Color online) (a) Typical aspect of the logarithm of the cost function $F(W,D,\alpha_0)$ for a fixed value of $\alpha_0$. The logarithm is shown here in order to enhance the visual contrast. This cost function has one minimum denoted ($W^*, D^*$) that depends on $\alpha_0$. (b) Typical aspect of the function $\log\left(F(W^*,D^*,\alpha_0)\right)$ for varying values of $\alpha_0$ ($W^*$ and $D^*$ being calculated for each $\alpha_0$). This cost function has a single minimum.

\item[] Figure 2. (Color online) (a) Backscatter coefficients for blood sheared at different residual shear rates and measured with the 0$\%$ SC concentration phantom (no attenuation), and corresponding fitting with the classical SFSE with no compensation for attenuation. (b) Backscatter coefficients for blood sheared at 5 s$^{-1}$ and measured with each of the four phantoms. The corresponding fitted models are the SFSE for the 0$\%$ SC phantom, and the SFSAE for the three other skin-mimicking phantoms (0.25, 0.5 and 0.75$\%$ SC).

\item[] Figure 3. (Color online) (a) Values of $W$, $D$ and $\alpha_0$ (in dB/MHz) for different residual shear rates estimated by the classical SFSE for the 0$\%$ SC concentration and by the SFSAE for the three skin-mimicking phantoms. (b) Corresponding relative errors.

\item[] Figure 4. (Color online) Relative errors of the packing factor and aggregate diameter for the three skin-mimicking phantoms obtained with the SFSE (a) with no compensation for attenuation ($W_{nocomp}$ and $D_{nocomp}$), and (b) with attenuation-compensation using the attenuation values estimated in transmission ($W_{comp}$ and $D_{comp}$). Parameters $W_{nocomp}$ and $W_{comp}$ and similarly $D_{nocomp}$ and $D_{comp}$ are compared with $W_{ref}$ and $D_{ref}$, respectively.

\end{itemize}

\newpage
\thispagestyle{empty}

\begin{figure}
\begin{center}
\includegraphics[width=4.5in]{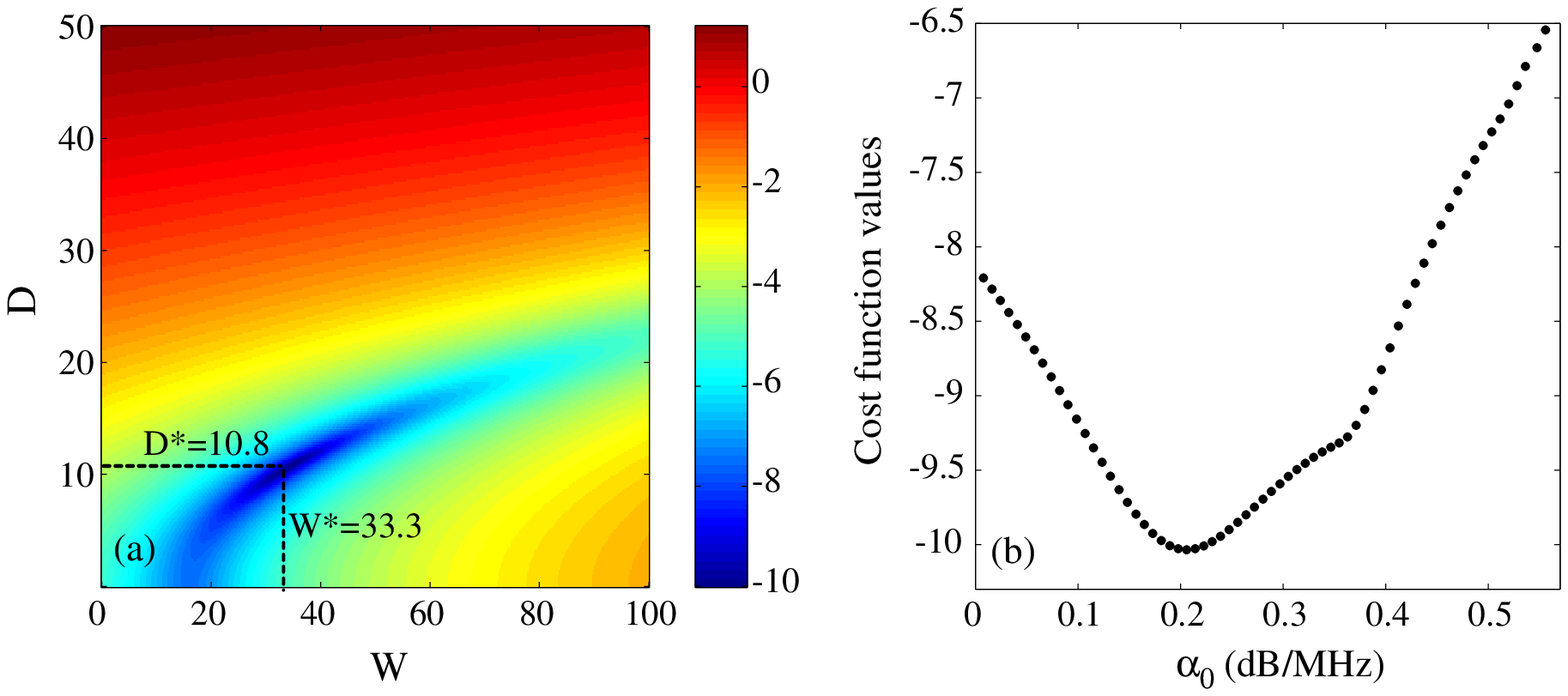}
\end{center}
\end{figure}

\vskip 2cm 
\begin{center}
Figure 1 
\end{center}

\newpage

\begin{figure}
\begin{center}
\includegraphics[width=5.5in]{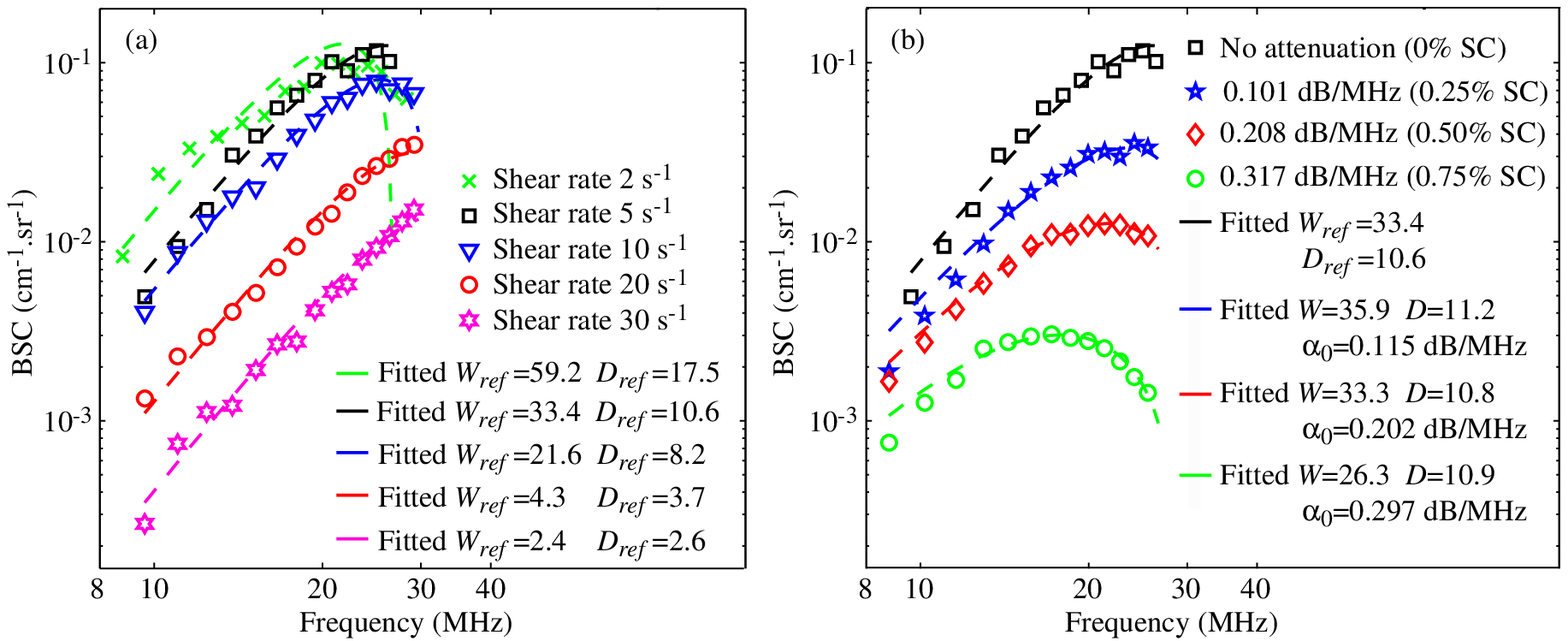}
\end{center}
\end{figure}

\vskip 2cm 
\begin{center}
Figure 2 
\end{center}

\newpage
\thispagestyle{empty}

\begin{figure}
\begin{center}
\includegraphics[width=4.5in]{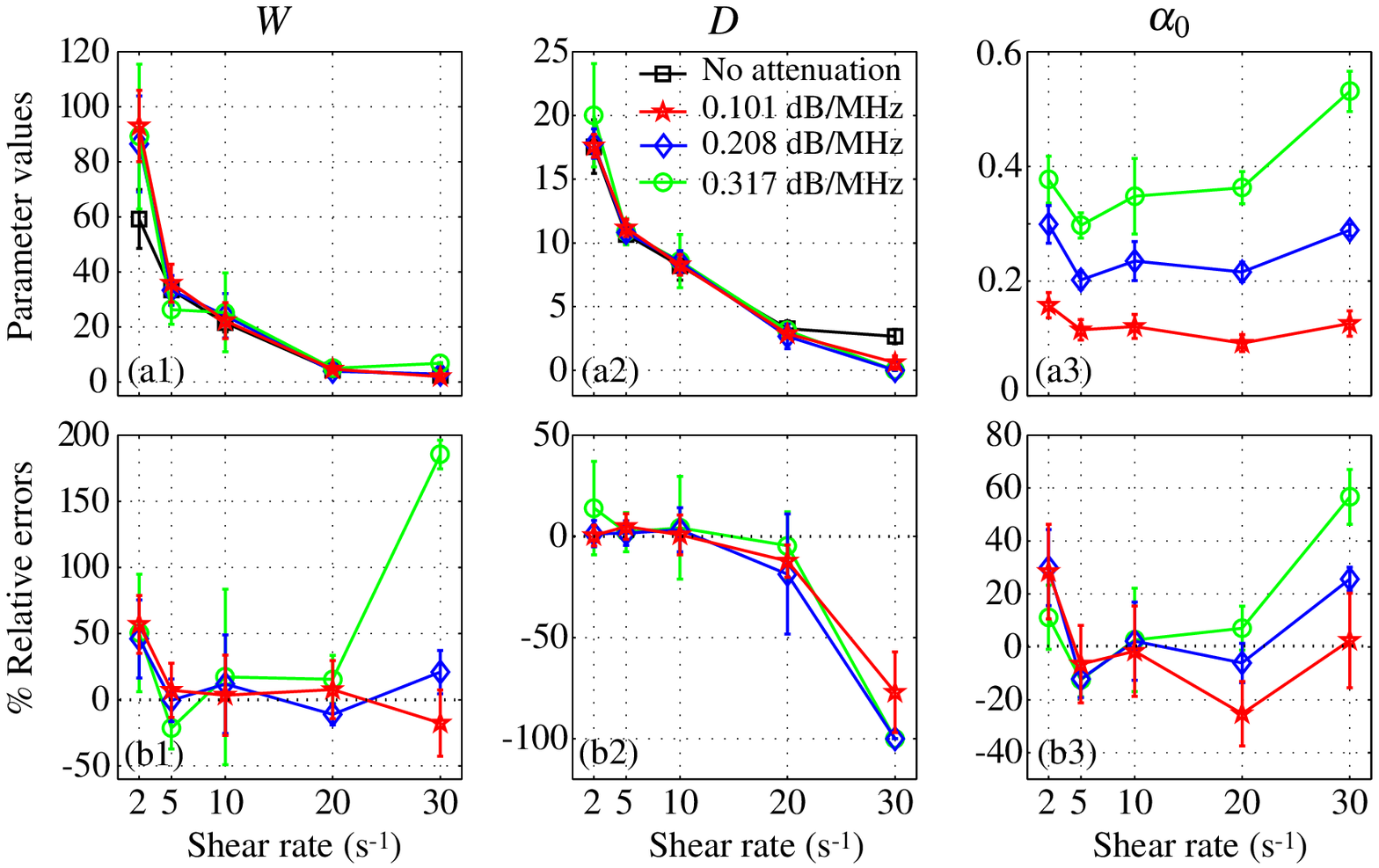}
\end{center}
\end{figure}

\vskip 2cm 
\begin{center}
Figure 3
\end{center}
 
\newpage
\thispagestyle{empty}
\begin{figure}
\begin{center}
\includegraphics[width=3.3in]{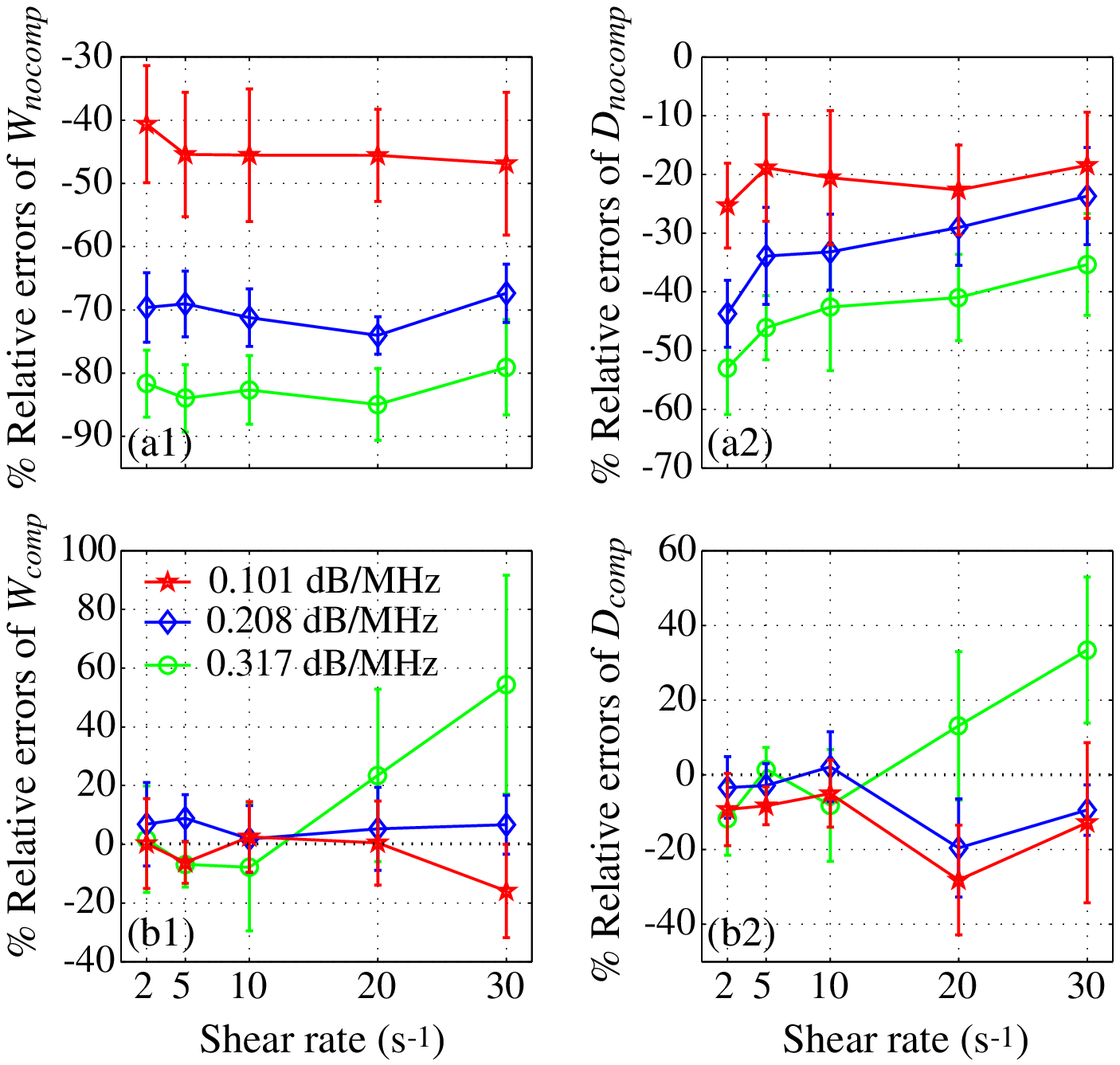}
\end{center}
\end{figure}

\vskip 2cm 
\begin{center}
Figure 4
\end{center}

\end{document}